\documentclass[mmnp]{edpsmath}
\usepackage{graphicx}
\usepackage{color}
\begin{document}
\title{Hybrid Modelling in Oncology:\\ Successes, Challenges and Hopes}\thanks{IDEX International Strategic Partnership Grant 2017, Universit\'e Grenoble Alpes}

\author{Ang\'elique St\'ephanou}\address{Universit\'e Grenoble Alpes, CNRS, TIMC-IMAG/DyCTIM2, 38041 Grenoble, France}
\author{Pascal Ballet}\address{University of Brest - LaTIM, UFR M\'edecine - IBRBS, 29238 Brest Cedex 3, France}
\author{Gibin Powathil}\address{Department of Mathematics, Computational Foundry, College of Science, Swansea University, Swansea SA1 8EN, United Kingdom}


\begin{abstract} 
In this review we make the statement that hybrid models in oncology are required as a mean for enhanced data integration. In the context of systems oncology, experimental and clinical data need to be at the heart of the models developments from conception to validation to ensure a relevant use of the models in the clinical context. The main applications pursued are to improve diagnosis and to optimize therapies.We first present the \textit{Successes} achieved thanks to hybrid modelling approaches to advance knowledge, treatments or drug discovery. Then we present the \textit{Challenges} than need to be addressed to allow for a better integration of the model parts and of the data into the models. And Finally, the \textit{Hopes} with a focus towards making personalised medicine a reality.    
\end{abstract}

\runningtitle{Hybrid Modelling in Oncology}



\maketitle

\section{Introduction}

Mathematical modelling in cancer is not new \cite{Laird1964}, however this fields truly exploded at the turn of the millennium. But at that time, the immense majority of models aimed to decipher, highlights or explain some mechanisms in relation to tumour growth, angiogenesis and invasion \cite{Chaplain1995}. The link with experimental data was scarce and essentially concerned the kinetics of growth. Progressively the models evolved in connection with two major advances: the first concerned the increased computational ability of the machines that allowed to perform simulations in the sense of numerical experimentations; the second concerned the progresses made in imaging techniques that allowed a wider access to data. With time, the models progressively became more "informed" meaning that they integrated experimentally measured parameters and were validated based on experimental (\textit{in vivo} or \textit{in vitro}) observations. Whereas the model developers were mainly located in mathematical, computing or engineering departments, it is now not rare to found them at the heart of hospital infrastructures.\\ 

As of today, models are massively used in the domain of cancer to pursue four main goals: to improve diagnosis, to improve therapy, to identify and develop new drugs and to bring new knowledge on the development of the disease. Those goals contribute to bring the models much closer to the clinic. Models are developed in the context of Systems Oncology \cite{Powathil2015} which - as Systems Biology \cite{Stephanou2018} - provides the comprehensive framework in which cancer can be investigated to truly understand and link its many aspects from the depth of the genes to the host environment. Such an integrated approach requires hybrid models, \textit{i.e.} models made of different sub-parts heterogeneously defined mathematically to adequately describe the events occurring at the many different spatio-temporal scales. However rather than focusing only on the hybrid aspect of the mathematical model formulation, we would like to extend the hybrid definition to the integration of experimental data at the core of the models. This crucial integration is more and more emphasized \cite{Anderson2018a,Jarrett2018,Ji2017a,Yankeelov2016a,Yankeelov2014}. As for multiscale integration, data integration requires specific model adaptation to accommodate heterogeneous information and/or indirect information in the sense that the required ones are often unavailable \cite{Jarrett2018}.\\

This review paper on hybrid modelling in oncology is organized in three parts. First, we present the \textit{Successes} with some examples where mathematical modelling in general and hybrid modelling in particular proved useful in a clinical context. Then we will evoke the \textit{Challenges} that computational biologists and biomathematicians have to face to fully exploit the huge potential of modelling so as to make the models commonly used in the clinical practice to diagnose and treat patients. Finally we will describe the \textit{Hopes} that the new generation of hybrid models bear towards making personalised medicine a reality.

\section{The Successes}

Mathematical oncology has now a proven track record : to understand, to predict, to optimize and to discover \cite{Anderson2018a}. In this section, we present some few representative contexts in which  mathematical models proved particularly insightful and/or useful.

\subsection{Prediction of cancer evolution}

The first models in cancer essentially aimed at describing the growth kinetics of the tumours \cite{Laird1964,Looney1975}. Beyond the simple description of growth, the models then developed to account for more detailed information such as the shape of the tumour, the state of the tumour cells (proliferative, quiescent/dormant or necrotic), the nature and state of the environment (vascularized, degraded, fibrous, hypoxic, acidic, \textit{etc.}) and also to predict how all these elements evolve when therapy comes into play \cite{Jarrett2018}. As for now, it is difficult to find clinical studies for which therapies have been defined or optimized from a mathematical model. It is most usual and widespread to find models that retrospectively predict how the tumour would grow or explain why a therapy was successful or failed. The obvious reason is that such experimentations on patients cannot be authorized since the risk remains high and clinicians are not yet ready to trust a computer program. In the other hand, models are now accepted in the clinic to predict the evolution of a tumour from patient images, to inform the clinician on the potential outcome. One such example is given in Colin \textit{et al.},\cite{Colin2014}, where temporal CT scans of a specific patient are used to calibrate a mathematical model that proved successful in predicting the evolution of metastatic nodules in the lung. In a similar way the study by Wang \textit{et al.},\cite{Wang2009} exploits magnetic resonance images (MRI) of patients with glioblastoma to measure for each individual patient the net cell proliferation and invasive rates. Those rates are then used in a mathematical model to evaluate the prognostic (survival time) for each patient. It has to be noted that these two cases are both excellent examples of hybrid models that integrate information from clinical images at their core. 

\subsection{Therapeutic optimization}

The number of mathematical models dedicated to therapeutic optimization is enormous, however computer-aided treatments at the bedside remain extremely limited as noted in \cite{Barbolosi2016}. The work accomplished towards treatment optimization clearly paves the path to personalised medicine where treatments would be designed to suit a specific patient by integrating all his/her characteristic from genetic to personal habits \cite{Stephanou2018}. We will not enter here into model presentations since this will be the main focus on the "\textit{Hopes}" section. One interesting example though is the case of chronotherapy whereby drug administration is adjusted to the circadian rhythms \cite{Ballesta2017a}. This example is worth mentioning since clinical trials showed an existing benefits of chronotherapy for cancer patients outcomes. At the same time, inter-patient variability was also exhibited. These findings advocate the necessity of personalised chronotherapeutics through integrated models. Those are thoroughly reviewed in Ballesta \textit{et al.},\cite{Ballesta2017a}.

\subsection{Identification of resistance to therapy}

One important contribution of the mathematical models was in the identification of resistance mechanisms. Passive resistance mechanisms for example could be involuntary exhibited in simulations while testing drug delivery regimens. In these simulations, it occurred that the drug could bypass the tumour targets depending on the structure of the vascular network \cite{Stephanou2005}. More generally, modelling of the spatiotemporal availability of drugs commonly show heterogeneous distribution in the tumour where portion of it might escape to treatment. Similarly, and for the same reason of non optimal oxygen delivery, part of the tumour can become hypoxic and the cell can enter into quiescence as a survival mean \cite{Stephanou2017}. Through this mechanism, the cells can escape to cytotoxic drugs since they stopped their cycle and limit the exposition to cytotoxic damages\cite{Pons-Salort2012}. Active resistance mechanisms have also been extensively modelled. Those mainly relate to the genetic and epigenetic alterations that modify the cells vulnerability to the drugs and/or enhance its evasion power. We refer the interested reader to the dedicated review by Foo \textit{et al.},\cite{Foo2014}.

\subsection{Identification of new drugs}

The new drug candidates are first tested \textit{in vitro} on cultured tumour cells where viability \textit{versus} death is measured in time for a range of drug concentrations. The dose is then scaled up for \textit{in vivo} testing on animals (mostly rodents) bearing the same tumour cells of interest. Finally clinical trials allow to test the drug on patients to fully evaluate the drug efficacy as well as its potential side effects. At the end of this long process, only a very small number of drug are finally approved by the various agencies \cite{Kola2004,Hutchinson2011} and even so the efficacy often remain below expectations \cite{Davis2017}. Mathematical modelling thus appear as a crucial tool for integrating \textit{in vitro} and \textit{in vivo} and make sense of the results so as to favour a more informed and efficient transposition to patients \cite{Carrara2017d}. PK/PD models are commonly used and accepted in drug development to help in selecting the best administration schedule, to avoid toxicity and to "unable decision making" \cite{Garralda2017}. However standard PK/PD are often limited on their own since other crucial information might be needed to catch all aspects of the drugs interaction with the overall system. In the very recent years PK/PD models were found to be coupled to other model elements such as the cell cycle, the circadian rhythms \cite{Ballesta2014}, or the cell physiology which led to PBPK-PD models (Physiologically-Based PK-PD) \cite{Ballesta2014,Block2015,Carrara2017d}. These couplings with detailed mechanistic models make it possible to identify new therapeutic targets and to help design new innovative drugs \cite{Carrara2017d}.

\section{The Challenges}
\subsection{Integration of experimental and mathematical models}

A mathematical or computational model is by essence always wrong in the sense that it can never entirely describe and represent all aspects of reality. It is always based on assumptions made by the modeller and is therefore strongly biased by the choices made. However this does not prevent models to produce accurate prediction on some aspects of the reality. For example, the speed with which a tumour is growing, its shape or its invasiveness can all be predicted based on some sets of relatively simple assumptions regarding proliferation rate or cell-cell adhesion \cite{Gerlee2009a}. Similarly, and as we saw it in the previous section, the effects of different types of therapy can also be assessed with a good accuracy and can include the prediction of resistance \cite{Foo2014,Hamis2018} or relapse effects \cite{Kim2015,Gallaher2018}. Phenomenological models - that is models that implement a predetermined answer for a phenomena given an entry condition or stimuli - often proved sufficient to make good qualitative predictions. For example to model the effect of a drug, it can simply be assumed that if the cells are subjected to a threshold concentration of drugs then it will enter apoptosis \cite{Pons-Salort2012}. In the other hand the explicit description of the mode of action of the drug can be modelled through mechanistic pharmacodynamic models \cite{Ballesta2014b}. This option is favoured to understand how the drug truly works and how its effect can potentially be enhanced by identifying some best suited conditions of administration or applicability. Phenomenologic or mechanistic, it is the utility and power of the model to answer the specific questions for which it has been built that matters the most. However in the context of cancer, and more specifically in the domain of drug testing, it is essential to integrate biological data at the heart of the computational model development to guide the choice of pertinent hypotheses and ensure the model validity to describe key observations. We indeed recall that it is illusory to search for model genericity - since each model serves its own purpose - but in the other hand it should be tailored to provide real insights to truly advance knowledge and to be immediately useful especially towards personalised medicine.\\

There is indeed a critical need for models truly rooted in reality (\textit{i}) so as to be relevant for a clinical use, (\textit{ii}) to cope with the high rate of failed clinical trials in the last remaining stages of drug testing \cite{Kola2004}. The enhanced integration between pre-clinical models (\textit{in vitro}  and \textit{in vivo}) with more informed mathematical models (\textit{in silico}) would lead to a much higher level of confidence, however efforts in that direction are still rare. This coupled \textit{in vitro/in vivo/in silico} approach was already attempted by \cite{Yankeelov2012} back in 2012 and was further advocated in \cite{Block2015} with the example of Physiologically-based pharmacokinetic and pharmacodynamic models (PBPK/PD).

\subsection{Management of the rising computational cost}

Hybrid modelling directed towards simulating the rise and development of cancer requests high computational power. What is generally costly is:
\begin{itemize}
\item the integration of the many interacting phenomena occurring at different time and space scales. This requires specific algorithms to order the simulations for the different events sequentially \cite{Pons-Salort2012,Lesart2012}. Time can be wasted since some events are on hold while other needs to reach stationary states for example. Moreover each event is generally solved with its own grid size and granularity from continuous to discrete \cite{Glade2013}. This is however amenable to parallel computing.
\item the simulation of a high number of cells, up to the million, since most models now describe the real 3D geometrical context and environment to account more faithfully to reality by avoiding geometrical biased \cite{Grogan2017}.
\item the integration of the tumour environment including other cell types (fibroblasts, immune or stem cells, etc.), the extracellular matrix and the tissue vasculature. This requires multiphysics modelling to account for the mechanical properties of the matrix or the blood flow through the vessels.
\item the graphical representation of the running simulation. This aspect can be very useful especially in the conception phase of the code. It allows to see the developing tumour and to image the variables that characterize the environment so as to see if they are correctly integrated. This also allows to stop the simulation - while testing some parameters for example - if inappropriate or deviant behaviours occur. 
\item the integration of experimental data. Since, as we saw it in the previous section, the new generation of hybrid models is to merge pre-clinical data in the \textit{in silico} model in a dynamic way. For example the data can serve as check points during the simulations (for validation) or can be provided and processed iteratively (scan images for example) so as to adapt a target trajectory for treatment optimality.
\end{itemize}
Of course not all models need to integrate all of these items, but the most recent hybrid models usually integrate at least one of them and each are computationally demanding on their own. It has to be noted that for most of these items there exists dedicated software, to solve multiphysics problems (e.g. Comsol multiphysics), mechanical issues (\textit{e.g.} Ansys), to deal with image processing on the fly (e.g. ImageJ), to graphically represents the variables while simulating (e.g. Paraview). Interfacing such different softwares could appear useful but in practice this can be very complicated not only for the implementation, but to maintain the workability of the overall code since it is required to cope with the many updates and releases of the individual softwares. Those can compromise their compatibilities and threaten the code sustainability on the long run. Moreover calling for multiple sub-codes and softwares generally takes too much machine time (RAM). This option is therefore not adapted and we are left with the necessity of recoding everything in an integrated and consistent framework.\\

\noindent\textit{Some solutions}\\
Although there is room to produce optimized algorithms, the best hope to cope with the rising computational costs involved in hybrid cancer modelling is to exploit the progress made in computer hardwares and processors. The computing power of the machines - that continues to grow exponentially - have been significantly enhanced over the last ten years with the appearance of the first multicore processors. Those allowed to overcome the stabilization (below 5GHz) of the clock speed of the single processor \cite{Ballet2018}. However, the graphics cards have today the highest computing power thanks to their massively parallel processor architecture of up to a thousand calculus units. They are programmable thanks to general purpose language close the C language like Cuda or OpenCL and can be used to compute parallel-structured algorithms. Quantum computing could be the next breakthrough to drastically improve the computing speed, although the applicability mostly concerns combinatory calculus \cite{Ballet2018}.\\

Coping with the simulation time is a real issue to address since the all goal of cancer modelling is to help in providing a faster and more accurate diagnosis and/or to help in defining the optimum treatment strategy. The models are thus designed to be used in the clinical context and as such - once fed with the data from the patient - they should provide an (almost) immediate answer to be approved as a clinical assistant.

\subsection{Need for software tools for research and education}

Given the increased complexity of the hybrid models, advanced computational skills become more and more essential in particular for those who wish to fully exploit the new hardware architectures. However, the main interest for a biomathematician or computational biologist is to focus on the modelling activity which is to define some work hypotheses to put under the test by identifying the key phenomena and the main actors (variables) of the subject of study. The time spent on the implementation of the model, \textit{i.e.} on coding, should not overcome the time spent on developing the model.\\

Fortunately many efforts have been made over the last 10 years to develop integrated computing environment that are open-source and free so as to be re-used and shared. Those environment dedicated to multicellular modelling all allow to implement easily basic cell properties, such as migration, adhesion and proliferation as well as reaction-diffusion dynamics of chemicals. We here focus on the presentation of a few of them among those which proved particularly useful to model tumour growth and the effects of therapies and whose used is widespread and acknowledged in the cancer modelling community. All are suitable for multiscale modelling, they allow to make 3D simulations and they all integrate a graphic interface to display the simulations.\\

\noindent\textit{CompuCell3D}\footnote{http://www.compucell3d.org/}\cite{Swat2012}\\
This software is based on the Cellular Potts Model (CPM) \cite{Graner1992}. It allows to describe the evolution of cells, where each cell is defined by a collection of sub-elements of a square grid. An energy function (Hamiltonian) that includes adhesion energies and volume constraints is evaluated. If the energy is minimized the cell configuration can be changed. Add-ons on the main software allow to integrate the intracellular dynamics \cite{Andasari2012}. Model implementation is flexible and can easily be performed by non-experts in coding by defining the model in an XML file. More experts users can produce their own code using the Python language and developers can even create their own plugins using the C language \cite{Ballet2018}. This framework has been extensively used to study angiogenesis \cite{Palm2016} but also to study therapy issues such as the routes of drug delivery \cite{Winner2016} or bystander effects in radiation therapy \cite{Powathil2016}.\\

\noindent\textit{Cell-based Chaste}\footnote{http://www.cs.ox.ac.uk/chaste/}\cite{Mirams2013,Pitt-Francis2009}\\
This is an efficient software in the sense that virtually anything can be modelled from regulation networks, to mechanical constraints. The implementation is based on three interlinked modules: (1) a cellular behaviour module allowing the intrinsic cell evolution (through its cycle for example), (2) a module for cell movement and mechanical interactions, (3) a module for the transport of substances (molecules for example). A recent useful add-on is specifically dedicated to the modelling of microvessels \cite{Grogan2017}. The strength of this software lies in its flexibility allowing lattice-free implementations as well as lattice-based approaches including the CPM. Its use however requires to possess more computing skills than for CompuCell3D but in the other hand it is well adapted to merge to the platform some pre-existing codes which can favour the adhesion of a wider modelling community. Although Cell-based Chaste was initially developed to model colorectal cancers \cite{vanLeeuwen2009}, it can potentially be applied to any type of cancers including non-solid tumors.\\

\noindent\textit{PhysiCell}\footnote{http://physicell.mathcancer.org/}\cite{Ghaffarizadeh2018}\\
PhysiCell stands for physics-based multicellular simulator. It is an agent-based simulator written in C++ and parallelized with OpenMP so as to simulate up to a million of cells. It integrates different solvers to describe the biochemical environment (from diffusing substrates to cell-secreted signals) as well as the cell-cell mechanical interactions. Paraview is chosen for data visualization. It is still a young platform but conceived to evolve with the emerging new needs. One strength of the platform is that it is sufficiently flexible for replicating results from other simulators which - as chaste - favours the migration of existing codes to this platform. Another advantage of the software is to put forward the possibility of a direct integration of experimental data file by using annotated experimental or clinical image as inputs to the software (although not implemented yet).\\

\noindent\textit{SimCells}\footnote{http://virtulab.univ-brest.fr/simcells.html}\cite{Ballet2018}\\
This software has been specifically developed to exploit the multiprocessor architecture of GPU. The number of simulated cells thus depends on the number of units of the GPU. It can reach the million with the most recent graphic cards. The cells in SimCells are defined as off-grid agents. A graphical user interface enables the user to define its systems with cells of different types whose behaviours are determined through the parametrization of predefined sets of rules and/or properties (such as adhesivity or motility) under the form of \textit{Conditions then Actions}. One advantage of the software is to give the ability to stop and restart the simulations so as to modify some parameters or conditions on the fly. The software is thus very useful to rapidly test some simple ideas without requiring any coding expertise. It is consequently an excellent tool for non expert in coding and proved particularly useful as a pedagogic tool to initiate a wide (diverse) audience to model development. The main drawback of the software for now is that it does not allow to represent the intrinsic physical/mechanical properties of the simulated objects. Moreover experimental data measurements can not be integrated since the definition of behaviours is essentially based on probabilities rather than on physical parameters, meaning that quantitative information cannot be reached. This is currently restraining the use of the software, however it remains a very useful numerical test bench,\textit{ i.e.} a good virtual platform for numerical bio-experimentations.

\section{The Hopes}
\subsection{Making personalised medicine a reality}

In a not too far future, the patient scans, analyses and medical history will feed a computer program. The MRI (or other scans) will be automatically analysed, the tumour volume evaluated, \textit{etc}. All information provided will be cross related and an accurate diagnosis would be made based on the currently developing deep learning algorithms where the computer will be trained to acquire a knowledge equivalent to this of a thousands of clinicians with different expertise altogether (radiologists, oncologists, neurologists, \textit{etc}.)\cite{Perez2018}. Beyond the diagnosis, the patient's data will be used to create the patient's virtual tumour \textit{i.e.} a numeric clone with the exact same location and properties (shape, volume, nature, grade, metabolic state, \textit{etc}.) so as to behave as the real tumour. Given the knowledge of the possible therapeutic means and of the usable drugs,  simulations would then been made using the virtual tumour to identify the optimum treatment in terms of means of action, dosage and planning of drug administration. Treatment adaptations during the course of treatment can be made by updating the virtual tumour with the newly available patient data. It is expected that such a clinical tool would produce at the same time a highly informed (hence accurate) diagnostic and the best adapted therapeutic strategy that would maximize the chances to cure the patient or to significantly extend the survival (by fully exploiting the current knowledge in oncology). Beyond the benefit to the patients, such a tool would help to standardize and optimize the screening procedures to make them more efficient so as to gain time to obtain a diagnosis and to start the treatment.\\

\begin{figure}[h!]
\centering 
\includegraphics[width=12cm]{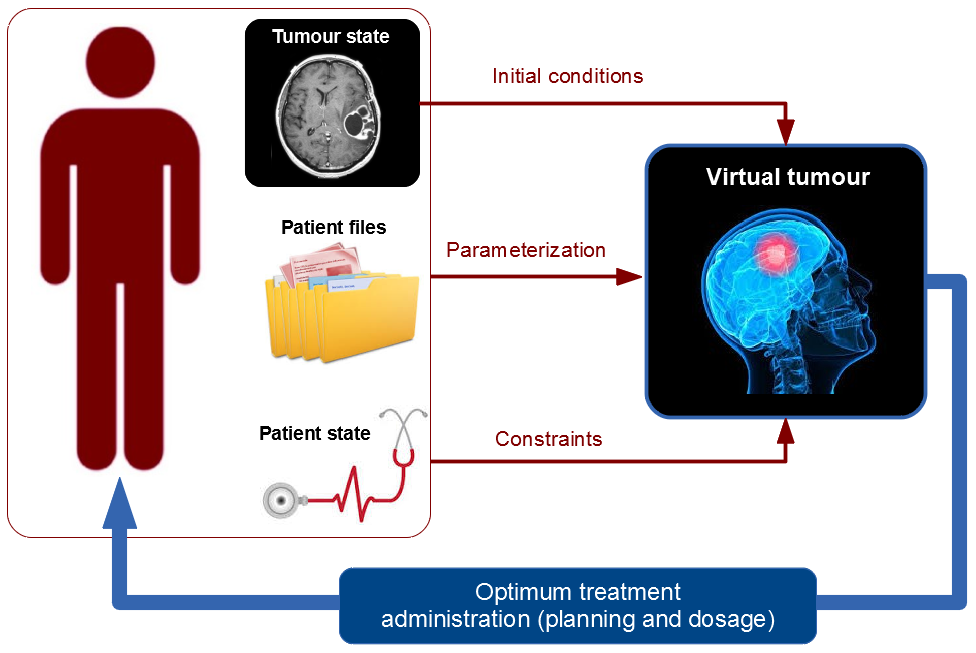}
\caption{The concept of personalised medicine. A virtual tumour is created from the patient data (scans/medical file/state) and is used to identify the optimum treatment and its schedule to cure or to maximize the life expectancy. All data acquired on the patient during the course of treatment is used to feed the virtual tumour and to update /adapt the therapy while respecting some imposed constraints (so as to control therapeutic interactions, to avoid contraindications with other pathologies, to take into account the patient state).}
\label{fig}
\end{figure}

The realization of such a tool requires a level of integration of highly heterogeneous biological data in multiple space and timescales, as illustrated in Figure \ref{fig}, making it a complex problem to address. However, current developments in computational, mathematical and systems oncology \cite{Caraguel2016, Karolak2018b, Barbolosi2016, Agur2014, Jackson2015, Baldock2013, Powathil2015, Stephanou2018, Poleszczuk2018, Enderling2013, Macklin2016} show great potential to develop predictive, personalised clinical cancer practice, integrating mathematical and computational approaches with traditional bench and clinical experiments. Moreover, the availability of vast amount of patient-specific data with the help of technical and computational advances can further help in developing such {\it in silico} framework/tool. Karolak {\it et. al} \cite{Karolak2018} gives an excellent review on various mathematical models that address tumour development, progression and response to treatments. Recently, there are several attempts and applications towards this direction, incorporating the complexity of tumour evolution and treatment efficacy in multiple scales \cite{Caraguel2016, Karolak2018b, Barbolosi2016, Jackson2015, Powathil2015, Stephanou2018, Poleszczuk2018, Macklin2016, Macklin2012}. Such models allow for intratumoural cross-scale integration of intracellular, extracellular and intercellular concepts, providing comprehensive modelling frameworks to which patient-specific information can be added to devise personalised therapeutic protocols.\\

Macklin {\it et. al.} \cite{Macklin2012, Macklin2016} introduced a patient-specific calibration method based on a hybrid agent based model to fully constrain the model based upon clinically-accessible histopathology data to obtain a patient histopathology providing new insights on the biophysical underpinnings of cancer. This may in future pave the way to augment a patient's imaging data with well calibrated models to predict optimal surgical margins based upon the patient-specific imaging data. Cook {\it et. al.} \cite{Cook2016} used a  biologically driven discrete hybrid cellular automaton (HCA) model of bone metastatic prostate cancer to obtain an optimal therapeutic window for putative targeted therapies as current pre-clinical models were limited in predicting the therapeutic effects. They used it to study the effect of TGF$\beta$ inhibitor treatment on the evolution of the cancer. St\'ephanou and co-workers \cite{Caraguel2016} proposed the design of a patient-specific virtual tumour incorporating angiogenesis, matrix remodelling, hypoxia, and cell state heterogeneity that will all influence the tumour growth kinetics and degree of tumour invasiveness. They have used a hybrid, multiscale approach for its implementation with the help of the pre-clinical data acquired on a mouse model as a proof of concept.  By combining image analysis and physiological modelling, they have shown that the simulated virtual tumour matches the characteristics and spatiotemporal evolution of the real {\it in vivo} tumour, showing the potential use of experimentally motivated computational and mathematical models in personalised therapeutic delivery. In a similar approach, Powathil {\it et. al.} \cite{Powathil2015, Brueningk2018} used a hybrid multiscale cellular automaton framework with intra- and inter-cellular dynamics to model tumour progression and further used to study the optimal combination of multimodal therapies. They have validated the model with experimentally estimated parameters for multiple cell-lines and their framework claims flexibility for modelling multimodality treatment combinations in different scenarios \cite{Brueningk2018}. Such experimentally calibrated and validated multiscale frameworks not only take us a step closer to personalised medicine but also helps as {\it in silico} test-bases for potential drug discovery and to study and test potential hypotheses to optimise multimodality therapeutic protocols before taking its clinical delivery.\\

In addition to such multiscale approaches, there are other mathematical models in the literature with a top-down approach to devise personalised cancer therapies. Swanson and co-workers \cite{Baldock2013,Massey2018,Jackson2015} use patient-specific mathematical models based on continuum mathematical modelling approaches to deliver patient-specific treatment predictions in neuro-oncology to treat very aggressive malignant brain tumours. Enderling lab  \cite{Poleszczuk2018,Prokopiou2015} developed mathematical model that use the concept of proliferation saturation index to predict patient specific radiotherapy protocols and they claimed that with the help of mathematical models, the information gained from radiobiological images can be used to select personalised RT dose-fractionation protocols. Ballesta {\it et. al} \cite{Ballesta2014} used a multiscale modelling approach, informed by {\it in vitro} to preclinical studies to develop cell line specific model to optimise anticancer therapies. Kronik {\it et.al} \cite{Kronik2010} proposed a general mathematical model for prostate cancer immunotherapy, incorporating vaccine interactions and immune system and validated the predictions using the results of a clinical trial data, showing its potential applications in personalised immunotherapy protocols.\\ 

These are some examples of several computational and modelling approaches to potentially deliver personalised multimodality therapeutic protocols in the fight against cancer. With the advancement of tissue engineering and the concept of "organ on chips" \cite{Esch2015,Perestrelo2015}, parameterisation, calibration and the validation of developing {\it in silico} tools can be achieved in a little more realistic scenarios and thus helping to move steps closer to providing patient-specific treatment delivery tools. 

\subsection{Deciphering the roots of the disease}

Beyond the interest for developing more efficient therapies, hybrid modelling in cancer can also prove essential to tackle more fundamental issues such as the questioning of the origin of cancer. An ongoing discussion concerns the Tissue Organization Fields Theory (TOFT) \cite{Soto2011} which is opposed to the classical Somatic Mutation Theory (SMT) \cite{Bedessem2015,Bizzarri2016,Bedessem2017}. The TOFT argues that a loss of tissue homeostasis consecutive to some damages in a tissue would favour the occurrence of mutations in the cells that found themselves coping with a state of environmental stress. In the other hand, the SMT is based on the mainstream understanding that mutations randomly occur. When they do not lead to cell apoptosis then more mutations can accumulate ultimately resulting in a deviant cell behaviour. If the latter is not spotted by the immune systems, then a cancer can develop.\\

Determining which theory is right can have enormous consequences in the way with which cancer is treated. In the SMT the cell is the cause and this justifies the use of most standard therapies that directly targets the tumour cells (cytostatic or cytotoxic molecules, radiation, surgery). In the other hand in the TOFT the cell is not the cause and the main focus should be to work on the environment and try to restore its normal (homeostatic) properties. This means that most therapeutic strategies need to be rethought accordingly to avoid a snow ball effect whereby therapy would be participating in a faster and more aggressive development of the disease.\\

Systems oncology and its hybrid modelling tools provide the means to fully investigate this delicate issue since it directly relates genotypic changes to tissue disregulation \cite{Basanta2017}. The question is: is it from bottom to top (SMT) or from top to bottom (TOFT)? As far as we know, no theoretical models (\textit{i.e.} simulation based) have been tacking this issue yet. Although we can imagine some ways to test this, using the same model hypotheses concerning differentiation and growth while testing alternatively (1) the effect of spontaneous mutations; (2) the loss of tissue homeostasis and its consequences on the cell behavioural divergence (phenotypic and genotypic). In both case, the evolution of the cells fitness (towards proliferation, migration or apoptosis) could be compared as a criterion for cancer progression.

\section{Conclusion}

We now enter the era of Artificial Intelligence where humans are assisted or replaced by machines for interpretational and decisional tasks that would lead to more accurate diagnosis and more efficient therapies. Without necessarily realizing it,  we - as biomathematicians, computational biologists, bioinformaticians and statisticians - are the main actors of this transition and evolution. The challenge for us is to build the theoretical tools that will make the machines able to process the information and produce interpretations that will condition decisions and actions. In this context of systems oncology, the new generation of models need to be more and more integrated to process both imaging data and "big data" (\textit{i.e.} genomics, protemics and other omics). Again, this calls for hybrid modelling to couple statistical models and deep learning algorithms for recognition and interpretation tasks with mechanistic models for knowledge-based predictions. These two modelling approaches - statistic and mechanistic - belong to two different scientific communities. One of the big challenge is to make them deeply interact so as to make personalised computer-assisted therapy a clinical reality.



\bibliographystyle{plain}
\bibliography{references}

\end{document}